\documentclass[prl,twocolumn]{revtex4}

\usepackage{graphicx}

\begin{document}

\title{Optical $\pi$ Phase Shift Created with a Single-Photon Pulse}

\author{Daniel Tiarks}
\author{Steffen Schmidt}
\author{Gerhard Rempe}
\author{Stephan D\"{u}rr}
\email{stephan.duerr@mpq.mpg.de}
\affiliation{Max-Planck-Institut f\"{u}r Quantenoptik, Hans-Kopfermann-Stra{\ss}e 1, 85748 Garching, Germany}

\begin{abstract}
A deterministic photon-photon quantum-logic gate is a long-standing goal. Building such a gate becomes possible if a light pulse containing only one photon imprints a phase shift of $\pi$ onto another light field. Here we experimentally demonstrate the generation of such a $\pi$ phase shift with a single-photon pulse. A first light pulse containing less than one photon on average is stored in an atomic gas. Rydberg blockade combined with electromagnetically induced transparency creates a phase shift for a second light pulse which propagates through the medium. Postselected on the detection of a retrieved photon from the first pulse, we measure a $\pi$ phase shift of the second pulse. This demonstrates a crucial step toward a photon-photon gate and offers a variety of applications in the field of quantum information processing.
\end{abstract}

\maketitle

\section*{INTRODUCTION}

Photons are interesting as carriers of quantum information because they hardly interact with their environment and can easily be transmitted over long distances. A deterministic photon-photon gate could be used as the central building block for universal quantum information processing (QIP) \cite{nielsen:00}. Such a gate can be built if a `control' light pulse containing only one photon imprints a $\pi$ phase shift onto a `target' light pulse \cite{Milburn:89}. As the interaction between optical photons in vacuum is extremely weak, an effective interaction between photons must be mediated by matter to create the required phase shift. Physical mechanisms that yield a large target phase shift created by a single control photon are difficult to find. One promising strategy is to couple an optical resonator to an atom, an atomic ensemble, or a quantum dot \cite{Turchette:95,Fushman:08,Parigi:12,Volz:14,Reiserer:14}. Another possible implementation is electromagnetically induced transparency (EIT) \cite{fleischhauer:05}. But for EIT with low-lying atomic states, the single-photon phase shifts measured to date are on the order of $10^{-5}$ rad \cite{Lo:11,Shiau:11,Feizpour:15}, which is much too small. It has been proposed to build a photon-photon gate by applying a small controlled phase shift to an ancillary coherent pulse with a large mean photon number \cite{Nemoto:04}. But the best performance achieved so far \cite{Feizpour:15} produced a phase shift of 18 $\mu$rad at a single-shot resolution for a measurement of the ancilla phase of 50 mrad, which is several orders of magnitude away from the goal. However, the combination with Rydberg states makes EIT very appealing \cite{Mohapatra:07,Gorshkov:11,Dudin:12,Firstenberg:13,Tiarks:14,Gorniaczyk:14}.

So far, three experiments \cite{Turchette:95,Fushman:08,Firstenberg:13} demonstrated an optical phase shift per photon between $\pi/10$ and $\pi/3$, only one experiment \cite{Volz:14} reached $\pi$. However, none of the schemes used there is applicable to deterministic optical QIP. Two of these experiments \cite{Firstenberg:13,Volz:14} measured self-phase modulation of a single continuous wave (CW) light field. But deterministic optical QIP requires that one light field controls another. The other two experiments \cite{Turchette:95,Fushman:08} measured cross-phase modulation (XPM) that one CW light field creates for another CW light field. However, an extension of these XPM experiments from CW light to a single photon, which must inherently be pulsed, is hampered by the fact that to spectrally resolve the two pulses, the pulses would need a duration exceeding the typical time scale of the XPM, given by the resonator decay time. Moreover, there is a no-go theorem \cite{Shapiro:06,Shapiro:07,Gea-Banacloche:10} which claims that it is impossible to achieve deterministic optical QIP based on a large single-photon XPM.

Here we show that the shortcomings of the existing experiments can be overcome by storing a control light pulse in a medium, letting a target light pulse propagate through the medium, and eventually retrieving the stored control excitation, similar to a proposal in Ref. \cite{Gorshkov:11}. Storage and retrieval circumvent the no-go theorem because that applies only to two simultaneously propagating light fields. We measure the controlled phase shift, that is by how much the presence of the control pulse changes the target pulse phase. We harvest the strong interactions in Rydberg EIT to create a large controlled phase shift. The incoming control pulse contains 0.6 photons on average. Postselected onto detection of a retrieved control photon, we obtain a controlled phase shift of $3.3\pm0.2$ rad.

\section*{RESULTS}

The experiment begins with the preparation of a cloud of typically $1.0\times10^5$ $^{87}$Rb atoms at a temperature of typically 0.5 $\mu$K in an optical dipole trap (see Materials and Methods) which creates a box-like potential along the $z$ axis, somewhat similar to Ref. \cite{Kuga:97}. We create Rydberg EIT with the beam geometry shown in Fig. 1A. The 780-nm signal beam propagates along the $z$ axis. This is an attenuated laser beam with Poissonian photon number distribution. The mean photon number in this beam is $\langle n_c\rangle= 0.6$ for the control pulse and $\langle n_t\rangle= 0.9$ for the target pulse. A 480-nm EIT-coupling beam used for the control pulse counter-propagates the signal beam. Another 480-nm EIT-coupling beam used for the target pulse co-propagates with the signal beam. The coupling light power $P_c$ is $(P_{c,c},P_{c,t})= (70,22)$ mW for control and target. The waists ($1/e^2$ radii of intensity) are $(w_s,w_{c,c},w_{c,t})= (8,21,12)$ $\mu$m. Using methods described in Ref. \cite{Baur:14}, we estimate coupling Rabi frequencies of $(\Omega_{c,c},\Omega_{c,t})/2\pi= (18,18)$ MHz. The coupling beams address principal quantum numbers $n_c=69$ and $n_t= 67$, see Fig. 1B. This pair of states features a F\"orster resonance with a van der Waals coefficient of $C_6= 2.3\times10^{23}$ atomic units \cite{Tiarks:14}. The timing sequence is shown in Fig. 1C.

\begin{figure}[!t]
\includegraphics[width=\columnwidth]{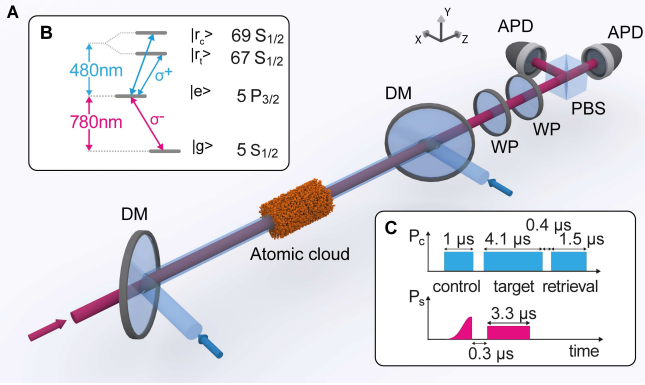}
\caption{\label{fig-scheme}
\textbf{Experimental Procedure} (\textbf{A}) Scheme of the experimental setup. Signal light (red) illuminates an atomic gas. Two additional beams (blue) provide EIT-coupling light, one co-propagating, the other counter-propagating the signal light. Dichroic mirrors (DMs) overlap and separate the beams. The polarization of transmitted signal light is measured using wave plates (WPs), a polarizing beam splitter (PBS), and avalanche photodiodes (APDs). (\textbf{B}) Level scheme. The 780-nm signal light couples states $|g\rangle= |5S_{1/2},F=2,m_F=-2\rangle$ and $|e\rangle= |5P_{3/2},F=3,m_F=-3\rangle$. The 480-nm EIT-coupling light couples states $|e\rangle$ and $|r\rangle= |nS_{1/2},F=2,m_F=-2\rangle$ with $n_c=69$ and $n_t= 67$ for control and target pulse. (\textbf{C}) Timing sequence of input powers $P_c$ and $P_s$ of coupling and signal light. The control pulse is stored in the medium, the target pulse propagates through the medium picking up a $\pi$ phase shift if a control excitation was stored, and eventually the control excitation is retrieved.
}
\end{figure}

Signal light transmitted through the atomic cloud is coupled into a single-mode optical fiber (omitted in Fig. 1A) to suppress stray light. After this fiber, the polarization of the light is measured using a polarizing beam splitter (PBS) and two avalanche photodiodes. The polarization measurement basis is selected using wave plates in front of the PBS. The probability of collecting and detecting a transmitted signal photon is 0.25.

\begin{figure}[!t]
\includegraphics[width=7cm]{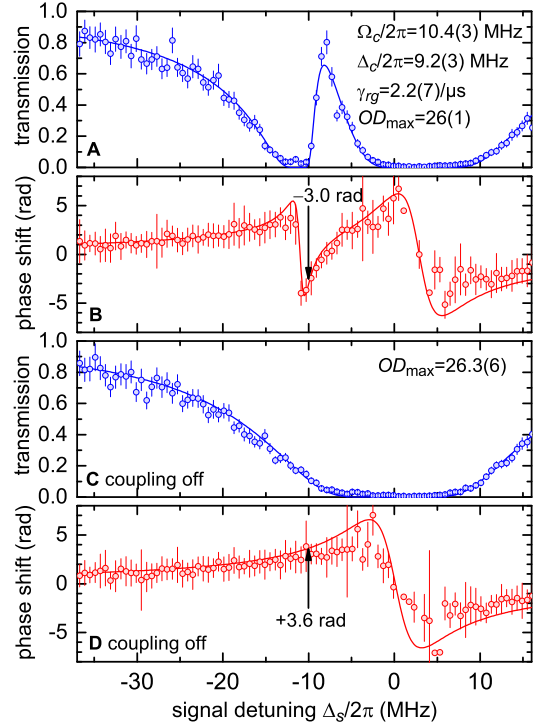}
\caption{\label{fig-spectra}
\textbf{Rydberg-EIT spectra without control pulse.} (\textbf{A},\textbf{B}) The transmission $e^{-OD_0}$ and phase shift $\varphi_0$ of the target signal beam are shown as a function of the signal detuning $\Delta_s$. The line in A shows a fit based on Eq. (3). The line in B shows the expectation from Eq. (3) for the parameter values obtained in A. For further experiments, we choose $\Delta_s/2\pi= -10$ MHz (arrow), which is near the minimum of $\varphi_0$. (\textbf{C},\textbf{D}) For reference, similar spectra are shown in the absence of coupling light. All error bars in this manuscript represent a statistical uncertainty of $\pm1$ standard deviation.}
\end{figure}

As shown in Fig. 1B, the target signal light is $\sigma^-$ polarized. To measure the phase shift that it experiences, we add a small $\sigma^+$ polarized component. This component serves as a phase reference because the phase shift that it experiences can be neglected since it is a factor of 15 smaller than for $\sigma^-$. Hence, a phase shift of the $\sigma^-$ component can be detected as a polarization rotation of transmitted target signal light (see Materials and Methods). Consider a target input polarization state $|\psi_{in}\rangle= c_+|\sigma^+\rangle+c_-|\sigma^-\rangle$, with amplitudes $c_+$ and $c_-$. Depending on whether 0 or 1 control excitations are stored, the output state will be
\begin{equation}
|\psi_{out,0}\rangle
\propto (c_+|\sigma^+\rangle+c_- e^{-OD_0/2}e^{i\varphi_0}|\sigma^-\rangle)\otimes|0\rangle
,\end{equation}
\begin{equation}
|\psi_{out,1}\rangle
\propto (c_+|\sigma^+\rangle+c_- e^{-OD_1/2}e^{i\varphi_1}|\sigma^-\rangle)\otimes|1\rangle
,\end{equation}
where OD$_j$, and $\varphi_j$ are the optical depth and the phase shift experienced by $|\sigma^-\rangle$ given that $j$ control excitations are stored. The goal is to achieve $\varphi_1-\varphi_0= \pi$.

Figs. 2A,B show measured EIT spectra of the transmission $e^{-OD_0}$ and phase shift $\varphi_0$ of signal light in the absence of the control pulse recorded with $1.0\times10^5$ atoms at a peak density of $\rho= 1.8\times10^{12}$ cm$^{-3}$. To model these quantities, we note that the electric susceptibility for EIT in a ladder-type level scheme, calculated analogously to Ref. \cite{fleischhauer:05}, is
\begin{equation}
\chi
= i \chi_0 \Gamma_e \left(\Gamma_e-2i\Delta_s+\frac{|\Omega_c|^2}{\gamma_{rg}-2i(\Delta_c+\Delta_s)}\right)^{-1}
,\end{equation}
where $\Gamma_e= 1/(26$ ns) is the population decay rate of state $|e\rangle$, $\gamma_{rg}$ the dephasing rate between $|g\rangle$ and $|r\rangle$, $\Omega_c$ the coupling Rabi frequency, $\Delta_s= \omega_s-\omega_{s,res}$ and $\Delta_c= \omega_c-\omega_{c,res}$ the single-photon detunings of signal and coupling light, $\chi_0= 2\rho|d_{eg}|^2/\epsilon_0\hbar\Gamma_e$ the value of $|\chi|$ for $\Omega_c= \Delta_s= 0$, $\epsilon_0$ the vacuum permittivity, $d_{eg}$ the electric dipole matrix element for the signal transition, and $\rho$ the atomic density. Propagating through a medium of length $L$ yields an optical depth of $OD= k_sL\mathrm{Im}(\chi)$ and a phase shift of $\varphi= k_sL\mathrm{Re}(\chi)/2$, where $k_s$ is the vacuum wave vector of the signal light. The best-fit values obtained from Fig. 2A agree fairly well with the expectations from the atomic density distribution, the coupling light intensity, and the value of $\gamma_{rg}$ measured at this density in Ref. \cite{Baur:14}. For later reference, Figs. 2C,D show spectra in the absence of coupling light.

We now turn to the effect that adding a control pulse has on $\varphi$. Note that unlike the target pulse, the control pulse is operated at $\Delta_s= \Delta_c= 0$ to optimize the storage efficiency. The combined efficiency for storage and retrieval of the control pulse is 0.2 for negligible delay between storage and retrieval and it drops to 0.07 after 4.5 $\mu$s in the absence of target light. The probability of storing more than one control excitation is suppressed by Rydberg blockade. Exploring what limits the efficiency is beyond the present scope. For dephasing due to thermal motion we expect a $1/e$ time of $\approx 30$ $\mu$s, indicating that other decoherence mechanisms dominate.

\begin{figure}[!t]
\includegraphics[width=7cm]{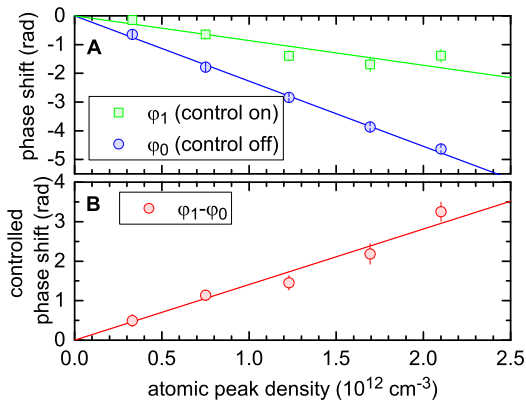}
\caption{\label{fig-density}
\textbf{Controlled phase shift.} (\textbf{A}) The phase shift $\varphi_0$ in the absence of a control pulse (blue circles) depends linearly on atomic density. So does the phase shift $\varphi_1$ in the presence of a control pulse (green squares) but with a different slope because of Rydberg blockade. (\textbf{B}) The difference between the two phase shifts yields the controlled phase shift $\varphi_1-\varphi_0$, which equals $3.3\pm0.2$ rad for the rightmost data point. The lines show linear fits.}
\end{figure}

Figure 3 shows a measurement of the controlled phase shift $\varphi_1-\varphi_0$ at $\Delta_s/2\pi= -10$ MHz. Clearly, a controlled phase shift of $\pi$ is reached. The $\varphi_1$ data (green) were postselected on detection of a retrieved control excitation to eliminate artifacts from imperfect storage efficiency. To change the density, the atom number loaded into the trap was varied, which had little effect on the atomic temperature. The first 0.8 $\mu$s of the target pulse are ignored in Fig. 3 because here transmission and phase show some transient, partly caused by the different group delays of $|\sigma^-\rangle$ and $|\sigma^+\rangle$.

There is a quantitative connection between Figs. 2 and 3. This is because when a control pulse is stored, Rydberg blockade pushes the EIT feature to very different frequencies (see Materials and Methods) so that, over the relevant frequency range, the blockaded part of the medium will take on the value of $\chi$ corresponding to Fig. 2D. If one excitation blockaded the complete medium, then the reference spectrum measured in Fig. 2D would match $\varphi_1$. At $\Delta_s/2\pi= -10$ MHz, the lines in Figs. 2B,D would then predict a controlled phase shift of $\varphi_1-\varphi_0= 6.6$ rad. In our experiment, the blockade region has a length of $2R_b$, where the blockade radius \cite{Tiarks:14} is estimated to be $R_b= |C_6/\hbar\Delta_T|^{1/6}= 14$ $\mu$m, with the full width at half maximum (FWHM) of the EIT transmission feature $\Delta_T/2\pi= 3.7$ MHz extracted from Fig. 2A. With $L= 61$ $\mu$m (see Materials and Methods), we expect a controlled phase shift of $\varphi_1-\varphi_0= (2R_b/L) 6.6$ rad = 3.0 rad. This agrees fairly well with the measurement in Fig. 3B, where the linear fit displays a controlled phase shift of 2.5 rad at $\rho= 1.8\times10^{12}$ cm$^{-3}$.

For the fidelity achievable in a future quantum gate, it will be crucial how well the phase coherence between the $\sigma^+$ and $\sigma^-$ components of the target signal light is maintained when creating the controlled $\pi$ phase shift. This can be quantified in terms of the visibility $V$ (see Materials and Methods). For the rightmost data point in Fig. 3, the polarization tomography from which we extract $\varphi_1$ yields $V= 0.75\pm0.14$.

\section*{DISCUSSION}

To summarize, we implemented a scheme in which a control pulse containing one photon imprinted a phase shift onto a target light field and measured a controlled phase shift of 3.3 rad. Our implementation offers a realistic possibility to be extended to building a photon-photon quantum gate in a future application. As the $|\sigma^+\rangle$ polarization is no longer needed as a phase reference when operating the gate, the polarization state of the target signal pulse would immediately be available as one of the qubits. We emphasize that the phase coherence properties of this qubit have already been explored in our present work by measuring the visibility. The control qubit could be a dual-rail qubit \cite{Paredes-Barato:14,Khazali:15} consisting of two beams propagating parallel to each other in the same atomic cloud with a relative distance larger than the blockade radius, such that the target beam overlaps with only one of the rails. On the input and output side, the dual-rail qubit could conveniently be mapped onto a polarization qubit as in Ref. \cite{Choi:08}.

A related experiment was simultaneously performed at MIT \cite{PiPhase:Beck:1512.02166}.

\section*{MATERIALS AND METHODS}

\subsection*{Optical Dipole Trap}

The dipole trap consists of a horizontal laser beam with a wavelength of 1064 nm, a waist of 140 $\mu$m, and a power of 3.2 W. This causes negligible axial confinement along the $z$ axis and a measured radial trapping frequency of 90 Hz. In addition, we use two light beams with a wavelength of 532 nm, waists of 25 $\mu$m, and powers of 0.10 W each. These `plug' beams perpendicularly intersect the dipole trapping beam and provide a box-like potential along the $z$ axis. The distance between the centers of the two plug beams is $\Delta z= 120$ $\mu$m. After carefully leveling the direction of the 1064 nm beam relative to gravity, the overall configuration creates a medium that is axially to a good approximation homogeneous. Using a polarizability of $\alpha= 711$ a.u. \cite{Marinescu:94a} for the $5S$ state at 1064 nm, where 1 atomic unit (a.u.) equals $1.649\times10^{-41}$ J(m/V)$^2$, we estimate a radial root-mean-square cloud size of $\sigma_r= 12$ $\mu$m. Using $\alpha= -250$ a.u. \cite{Saffman:10} for the $5S$ state at 532 nm, we estimate an axial FWHM cloud size of $L= 61$ $\mu$m. Note that $L$ is smaller than $\Delta z$ because the atomic temperature is much below the barrier height created by the plug beams. The radial inhomogeneity of the medium has little effect because $w_s< \sigma_r$. A magnetic field of $\approx$100 $\mu$T is applied along the $z$ axis to stabilize the orientation of the atomic spins. The two plug beams are generated by sending one light beam into an acousto-optic modulator driven with the sum of two sinusoidal radio frequency (rf) fields, thus generating two first-order diffracted output light beams. This makes $\Delta z$ and hence $L$ easily adjustable by changing the frequencies of the rf fields.

The dipole trap is loaded from a magnetic trap. Before the transfer, the atomic cloud is cigar shaped with the $x$ axis as the symmetry axis. After the transfer, the cloud is also cigar shaped but with the $z$ axis as the symmetry axis. This transfer from one elongated trap into another perpendicularly elongated trap is nontrivial. It turns out that the atom number fluctuations added by the transfer are minimized, if we first slowly ramp up another 1064 nm dipole-trapping beam propagating along the $y$ axis which forces the cloud into an almost spherical shape during transfer. Second, we slowly ramp up the dipole trapping beam along the $z$ axis together with the plug beams. Third, we slowly ramp down the magnetic trap and, finally, we slowly ramp down the dipole-trapping beam along the $y$ axis.

The 480-nm coupling light creates a repulsive potential which is added to the dipole trap potential. The coupling light is on for only few microseconds and the experiment is repeated every 100 $\mu$s. This low duty cycle is chosen because it makes the effect of the repulsive potential negligible, as in Refs. \cite{Tiarks:14,Baur:14}.

\subsection*{Polarization Tomography and Visibility}

To measure the phase shift $\varphi_j$ in Eqs.\ (1) and (2), we perform tomography of the output polarization state of the target signal light. To this end, the experiment is repeated many times for any given set of experimental parameters. In each repetition, one out of three polarization measurement bases is chosen. These bases are horizontal/vertical (H/V), diagonal/anti-diagonal (D/A), and left/right circular (L/R). Combination of these measurements yields the normalized Stokes parameters $S_{HV}$, $S_{DA}$, and $S_{LR}$, where $S_{kl}= (P_k-P_l)/(P_k+P_l)$ and $P_k$ is the light power in polarization $k$. The normalized Stokes parameters contain the complete information about the polarization state of the light. We express the normalized Stokes vector in spherical coordinates as $(S_{HV},S_{DA},S_{LR})= S_0(\sin\vartheta\cos\varphi, \sin\vartheta\sin\varphi, \cos\vartheta)$. Hence, we obtain radius $S_0$, polar angle $\vartheta$, and azimuth $\varphi$. We choose the amplitudes $c_+$ and $c_-$ of the input state of the signal polarization to be real and positive. Hence, the azimuth $\varphi$ equals, modulo $2\pi$, the phase shift $\varphi_j$ in Eqs.\ (1) and (2). Thus the tomography yields $\varphi_j$.

Alternatively, one could, in principle, determine the azimuth $\varphi$ by measuring in many bases with linear polarizations which include various angles $\alpha$ with the horizontal polarization. In this case, one would expect to measure a transmitted power $P_\alpha= P_\mathrm{total}[1+V\cos(\varphi-2\alpha)]/2$ with $P_\mathrm{total}= P_H+P_V$ and fringe visibility $V= S_0\sin\vartheta$.

From our polarization tomography measurements, we extract a visibility $V= \sqrt{S_{HV}^2+S_{DA}^2}$. The visibility characterizes the phase coherence properties of the polarization of the target signal pulse. The ratio of $c_+$ and $c_-$ in the input polarization was chosen to maximize $V$ for the measurement of $\varphi_1$.

\subsection*{Sign of the Rydberg-Blockade Shift}

If the propagating target excitation and the stored control excitation have a relative distance $r$, then their van der Waals potential is $V= -C_6/r^6$. For small $r$, this creates Rydberg blockade. In our experiment, the positive sign of $C_6$ implies an attractive van der Waals interaction which lowers the energy of the Rydberg pair state. At fixed detuning $\Delta_c$ of the EIT-coupling laser, the EIT feature is therefore shifted to smaller signal detuning $\Delta_s$, i.e.\ further to the left in Fig.\ \ref{fig-spectra}B. This is advantageous because there is no radius $r$ at which the left side of the EIT feature, where the phase shift is large and positive, would appear at $\Delta_s/2\pi= -10$ MHz. Hence, when continuously decreasing $r$ from infinity to near zero, $\mathrm{Re}(\chi)$ as a function of radius starts out at the value relevant for Fig.\ \ref{fig-spectra}B and monotonically approaches the value relevant for Fig.\ \ref{fig-spectra}D. This avoids the Raman resonance, as discussed in Ref.\ (14).

Had we reversed the signs of $\Delta_s$ and $\Delta_c$, then for decreasing $r$, $\mathrm{Re}(\chi)$ would first overshoot to large positive values at the Raman resonance before settling down to the value relevant for Fig.\ \ref{fig-spectra}D. Integration of $\mathrm{Re}(\chi)$ over distance would then yield a reduced controlled phase shift, which is undesirable. We tested this experimentally and found that the controlled phase shift was indeed reduced by a factor of roughly 1.5. A similar asymmetry under simultaneous sign reversal of $\Delta_s$ and $\Delta_c$ was observed in Ref. \cite{Firstenberg:13}.

\textbf{Acknowledgments:}
We thank Giovanni Girelli for experimental assistance during an early stage of the experiment. 
This work was supported by DFG via SFB 631 and via NIM.

\end{document}